\documentclass[11pt]{article}
\usepackage[margin=1in]{geometry}
\usepackage[T1]{fontenc}
\usepackage{lmodern}
\usepackage{microtype}
\usepackage{graphicx}
\usepackage{booktabs}
\usepackage{amsmath}
\usepackage{xcolor}
\usepackage[numbers,sort&compress]{natbib}
\usepackage[hidelinks]{hyperref}
\usepackage{enumitem}
\setlist{itemsep=2pt,topsep=4pt}

\newcommand{\TrialsPerAttack}{100}
\newcommand{\HandoffN}{1{,}000}
\newcommand{\BaseBlockedNetPolicy}{500 of 500}
\newcommand{\BaseIdentityAttacksSucceeded}{600}
\newcommand{\BaseIdentityAttacksTotal}{600}

\newcommand{\HardOverheadSentenceCap}{The added cost was small: for 1\,KB messages the relay added a median of 0.29\,ms per handoff over a direct route in the baseline and 0.53\,ms in the hardened variant, and hardening reduced relay throughput at 16 concurrent senders by 8\%.}
\newcommand{\HardBlockedSentence}{all 1{,}200 attempts}
\newcommand{\EnvHost}{2 vCPU, 8\,GB RAM, Linux 6.8.0-149-generic, Docker 29.1.3}
\newcommand{\DiscussionCost}{Mediation cost little per message: the paired median difference between relay and direct handoff was 0.29\,ms for 1\,KB messages in the baseline and 0.53\,ms in the hardened variant, so hardening added about 0.23\,ms per handoff (comparing paired differences reduces the effect of run-to-run drift, which moved the direct route itself by 0.24\,ms between runs). The dominant end-to-end cost was not mediation but polling: with the agents' 0.5\,s polling interval, median delivery took 265\,ms against 2.0\,ms for a direct send. A push-based relay would remove most of that delay without giving up mediation.}

\title{Can the Audit Trail Be Trusted?\\
Testing Clinical Governance Requirements\\
in a Contained Multi-Agent LLM Pipeline}

\author{
Anas H. Alzahrani, MD, MPH, PhD\thanks{Department of Preventive Medicine and Public Health, Faculty of Medicine, King Abdulaziz University, Jeddah, Saudi Arabia. Corresponding author: \href{mailto:ahalzhrani@kau.edu.sa}{ahalzhrani@kau.edu.sa}. ORCID: \href{https://orcid.org/0000-0002-1394-9157}{0000-0002-1394-9157}.}
\and
Omar F. Alaskar\thanks{College of Computer and Information Sciences, Imam Mohammad Ibn Saud Islamic University, Riyadh, Saudi Arabia.}
}
\date{}

\begin{document}
\maketitle

\begin{abstract}
\textbf{Background.} Clinical research groups are starting to build pipelines in which several large language model (LLM) agents work together, for example one defining a phenotype, another checking dosing rules and a third reviewing the result. Clinical governance requires that such a system keep each agent within its sanctioned role and keep a trustworthy record of which agent did what. A common engineering answer is containment: each agent runs in its own container on its own network, and all communication passes through a relay that checks a routing policy and writes an audit ledger. Whether containment delivers what governance needs has not been tested.
\textbf{Methods.} We translated four governance requirements (containment, attribution, non-interference and protected oversight) into twelve concrete attacks by a compromised agent, and ran each \TrialsPerAttack{} times against Connexin, an open containment system whose own verification suite had passed. We then built a hardened relay and repeated the attacks, and measured the latency and throughput cost of mediation and hardening.
\textbf{Results.} The baseline met the containment requirement, blocking \BaseBlockedNetPolicy{} attempts to bypass the relay or break routing policy. It failed the other three: the compromised agent succeeded in \BaseIdentityAttacksSucceeded{} of \BaseIdentityAttacksTotal{} attempts to impersonate another agent, read or silently consume another agent's messages, read the audit ledger, replay messages and reach the internet. Impersonated messages were written to the ledger under the forged sender, hijacked acknowledgements were logged as ordinary ones and reads of other agents' mailboxes left no record, so the audit trail looked authoritative but was wrong. The hardened relay, which takes identity from per-agent credentials rather than from message content, blocked \HardBlockedSentence{} and logged each refused relay request against the credential that actually made it. \HardOverheadSentenceCap{}
\textbf{Conclusions.} Containment alone does not produce an audit trail that clinical governance can rely on. Verifiable agent identity, owner-scoped access and protected audit data should be baseline requirements for multi-agent LLM systems used in clinical research, and they can be met at low cost.
\end{abstract}

\section{Introduction}
Clinical governance is the framework through which health organisations are accountable for the quality and safety of the care and research they deliver~\cite{scally1998clinical}. In practice it rests on knowing who did what: clinical audit, incident review and research integrity all depend on records that attribute each action to a responsible person or process. As LLMs enter clinical work, guidance from the World Health Organization calls for the same accountability over AI systems~\cite{who2024lmm}, and the EU AI Act requires high-risk AI systems to record events automatically over their lifetime so that their functioning can be traced~\cite{euaiact2024}.

These requirements become harder to meet when the AI is not one model but a team of agents. Multi-agent LLM systems divide a task among specialised agents that pass work to each other~\cite{wu2023autogen}, and the approach is being applied to medical decision-making~\cite{kim2024mdagents} and to clinical research pipelines, where one agent may define a cohort, another check exposure and dosing definitions and a third review the output against a guideline. Each agent may touch protected health information, and the combined output may shape a clinical conclusion. Governance then has to answer two questions about the pipeline: can any one agent act outside its role, and, after the fact, can we tell which agent produced which result?

Both questions matter because agents can be subverted. Content an agent reads can take control of it through indirect prompt injection~\cite{greshake2023indirect}, and in multi-agent systems an injected instruction can spread from agent to agent~\cite{lee2024promptinfection}. A governance-minded design should therefore assume that any single agent may become adversarial. The common engineering response is containment, borrowed from systems security: isolate each agent and mediate every interaction~\cite{wu2025isolategpt}. In practice this usually means containers on separate networks and a relay that enforces a routing policy and keeps a ledger, which looks like it provides both role enforcement and an audit trail.

We built such a system, Connexin, for clinical research pipelines, and its own verification suite passed. Those checks asked whether the system worked as designed. This paper asks whether it meets governance requirements when one of its agents turns against it.

\paragraph{Contributions.}
\begin{enumerate}
\item We translate four clinical governance requirements into testable properties of a multi-agent system and twelve concrete attacks by a compromised agent (\S\ref{sec:requirements}).
\item We show that a standard containment design meets the containment requirement but fails attribution, non-interference and protected oversight, and that its audit ledger records forged actions as genuine (\S\ref{sec:results}).
\item We describe a hardened relay that addresses all four requirements, and measure its latency and throughput cost (\S\ref{sec:hardening}, \S\ref{sec:results}).
\item We release the system, experiment kit and raw data, and give review questions for governance committees assessing multi-agent AI systems.
\end{enumerate}

\section{Governance requirements}\label{sec:requirements}
We derived four requirements from the accountability and record-keeping expectations above~\cite{scally1998clinical,who2024lmm,euaiact2024} and from the zero-trust principle that network location alone confers no trust~\cite{rose2020zerotrust}. Each is stated as what governance needs, followed by the property a system must have and the attacks we used to test it (Table~\ref{tab:attacks}).
\begin{description}[leftmargin=0pt]
\item[G1 Containment.] \emph{Each agent acts only within its sanctioned role.} An agent cannot reach other agents except through the mediator, and cannot send messages on routes or of types its role does not allow. Attacks: direct connections to other agents (A1, A2); routes, message types and receivers outside policy (A3, A4, A11).
\item[G2 Attribution.] \emph{Every action can be traced to the agent that took it.} The system must not accept, or record, an action under a different agent's identity, a replay of an earlier action, or an action from an unidentified caller. Attacks: impersonation (A5), replay (A9), unauthenticated requests (A12).
\item[G3 Non-interference.] \emph{No agent can read or alter another agent's work.} One agent must not be able to read the messages addressed to another, or mark them as handled so that they are never processed. Attacks: reading another agent's mailbox (A6), acknowledging its messages (A7).
\item[G4 Protected oversight.] \emph{Audit data belongs to governance, and data leaves only through sanctioned paths.} Agents must not read the audit ledger, which describes all agents' activity, and must not reach outside systems directly. Attacks: reading the ledger (A8), opening a connection to the internet (A10).
\end{description}
In every case the record of the attempt matters as much as whether it was blocked: governance needs refused actions to be logged against the agent that actually tried them.

\section{Related work}
\paragraph{Governance and accountability for clinical AI.} WHO guidance on large multi-modal models sets out recommendations for governments, developers and health-care providers on the accountable use of generative AI in health~\cite{who2024lmm}, and the EU AI Act makes automatic event logging a legal requirement for high-risk AI systems~\cite{euaiact2024}. Both assume that logs are faithful. We test that assumption for a multi-agent system under attack.

\paragraph{Containment of LLM agents.} IsolateGPT isolates third-party LLM applications in separate execution environments connected through a central hub that mediates every interaction~\cite{wu2025isolategpt}. Container isolation is itself imperfect: frontier models can exploit container misconfigurations and some runtime and kernel weaknesses to escape~\cite{marchand2026sandboxescape}. We do not test escape; we ask what an agent can do without escaping.

\paragraph{Access control for agents.} SEAgent frames agent attacks as privilege escalation, including a confused-deputy variant in multi-agent systems, and enforces attribute-based mandatory access control~\cite{ji2026seagent}. AgentGuard applies attribute-based access control to tool calls~\cite{luo2026agentguard}, and CaMeL separates control flow from data flow so that untrusted data cannot change what an agent does~\cite{debenedetti2025camel}. All three decide whether an action is allowed, and all depend on knowing which agent is acting.

\paragraph{Agent-to-agent communication.} The Agent2Agent (A2A) protocol standardises how agents discover each other and exchange tasks~\cite{a2a2025}. A security analysis found that A2A establishes identity only at the transport layer and does not carry it across delegation hops~\cite{lotfi2026a2abreak}, and a survey of open problems in multi-agent security lists authentication and accountability between agents among them~\cite{schroederdewitt2025openchallenges}. Our results measure what happens to an audit trail when identity is asserted rather than verified.

\section{System design}\label{sec:design}
Connexin runs each component as a separate container (Figure~\ref{fig:arch}).
\begin{itemize}
\item The \textbf{gateway} is the only component reachable from outside. It authenticates external callers with a bearer token and starts work by publishing to the relay.
\item The \textbf{relay} (message bus) is the only component attached to more than one network. Agents publish messages to it, and it checks each against a capability registry listing, for every agent, the receivers and message types it may use. Allowed messages are queued for the receiver; denied ones are rejected. Every decision is appended to a JSON-lines ledger with timestamps and policy-decision time.
\item \textbf{Agents} (here a \emph{worker} and a \emph{reviewer}) each sit on their own network, shared only with the relay. They retrieve messages by polling their mailbox on the relay every 0.5\,s and acknowledge each message they process.
\end{itemize}
Every container runs as a non-root user with all Linux capabilities dropped, \texttt{no-new-privileges}, a read-only root filesystem, and memory, CPU and process limits. The policy registry, ledger, metrics and a suite of verifiers for connectivity, isolation, ledger and policy were developed iteratively, and the full suite passed before this study.

\begin{figure}[t]
\centering
\fbox{\parbox{0.92\linewidth}{\centering\small
\texttt{external caller} $\rightarrow$ \textbf{gateway} \textit{[gateway\_net]}\\[2pt]
$\downarrow$\\[2pt]
\textbf{relay}: policy check $\cdot$ mailboxes $\cdot$ audit ledger \textit{[gateway\_net, alpha\_net, beta\_net]}\\[2pt]
$\swarrow\qquad\searrow$\\[2pt]
\textbf{worker} \textit{[alpha\_net]} \qquad\qquad \textbf{reviewer} \textit{[beta\_net]}\\[4pt]
\textit{No network is shared between two agents; all agent traffic passes through the relay.}
}}
\caption{Connexin baseline architecture. Each component is a hardened container; bracketed names are the Docker networks each is attached to.}
\label{fig:arch}
\end{figure}

\paragraph{Threat model.} One agent, the worker, is fully compromised, for example through prompt injection: the attacker controls its code and can send arbitrary requests from inside its container. The attacker has only what the worker legitimately has: its network attachment and, in the hardened variant, its own credential. For the acknowledgement attack we also give the attacker the identifiers of messages queued for another agent, a worst case for identifier leakage. The relay, container runtime and host are trusted, and container escape is out of scope.

\section{Hardened relay}\label{sec:hardening}
The hardened variant changes only what the requirements showed to be missing and leaves the policy registry, agents and container settings unchanged.
\begin{enumerate}
\item \textbf{Identity from credentials (G2).} Each agent receives its own random bearer token; only the relay holds the mapping from tokens to agent names. A publish request is rejected unless its token maps to the sender named in the message.
\item \textbf{Replay protection (G2).} Each request carries a nonce and a timestamp; the relay rejects requests outside a 30\,s window and any nonce already seen in that window.
\item \textbf{Owner-only mailboxes (G3).} Only the authenticated owner of a mailbox may read it or acknowledge its messages.
\item \textbf{Operator-only audit data (G4).} The ledger and metrics require the operator (gateway) credential.
\item \textbf{No internet route (G4).} All networks are created as internal networks, so containers have no route to the internet.
\end{enumerate}
Every refusal is written to the ledger with the caller's authenticated identity beside the identity it claimed. The change to the relay is about 80 lines of Python.

\section{Methods}\label{sec:methods}
\paragraph{Setup.} Both variants ran on the same host (\EnvHost) as separate, throwaway Docker Compose projects with no host ports, alongside other, unrelated workloads on that host. Images, resource limits and hardening flags were identical between variants apart from the changes in \S\ref{sec:hardening}. We kept the deployed resource limits: 1 CPU and 512\,MB for the relay, and 0.5 CPU and 256\,MB for every agent and for the direct receiver used in Experiment~2.

\paragraph{Experiment 1: governance attacks.} From inside the worker container we ran the twelve attacks in \S\ref{sec:requirements}, each \TrialsPerAttack{} times per variant. An attempt counted as \emph{succeeded} only if the attacker obtained what it wanted, for example a delivered forged message or the content of another agent's message. Before the attacks a legitimate sender queued \TrialsPerAttack{} canary messages for the gateway. Afterwards we audited the relay's ledger for every attempt: whether it was logged and to which agent it was attributed.

\paragraph{Experiment 2: cost of mediation and hardening.} Governance controls must be affordable to be adopted, so we measured what mediation and hardening cost. We compared delivering a message through the relay with sending it directly to a receiver on the sender's own network. The direct receiver used the same web framework and request model as the relay but performed no policy check and wrote no ledger, so the difference is the mediation itself. Senders used the agents' own HTTP client, opening a new connection per request.
\begin{itemize}
\item \emph{Handoff latency}: the time the sender waits for a message to be accepted. For 1, 10 and 100\,KB payloads we sent \HandoffN{} relay--direct pairs, randomising the order within each pair, after 20 warm-up pairs.
\item \emph{Concurrency}: 400 messages of 1\,KB shared among 1, 4 or 16 concurrent senders; we report throughput and per-request latency.
\item \emph{End-to-end delivery}: 200 messages sent at 10 per second, timed from sending to retrieval by the receiver, which polled every 0.5\,s (as the agents do) or every 0.05\,s; the direct receiver timed arrival.
\end{itemize}

\paragraph{Statistics.} For attack success we report counts with Wilson 95\% confidence intervals. For latency we report medians and 95th percentiles, with 95\% confidence intervals for medians and for paired relay-minus-direct differences from 2{,}000 bootstrap resamples.

\section{Results}\label{sec:results}
\subsection{Compromised agent}

\begin{table}[t]
\centering\small
\caption{Attempts in which the compromised worker got what it wanted, out of attempts made, grouped by the governance requirement each attack tests (\S\ref{sec:requirements}).}
\label{tab:attacks}
\begin{tabular}{lcc}
\toprule
Attack & Baseline & Hardened\\
\midrule
\addlinespace[2pt]\multicolumn{3}{l}{\textit{G1 Containment}}\\
A1\;Connect directly to the reviewer & 0/100 & 0/100\\
A2\;Connect directly to the gateway & 0/100 & 0/100\\
A3\;Publish on a route outside policy & 0/100 & 0/100\\
A4\;Publish a message type outside policy & 0/100 & 0/100\\
A11\;Publish to a non-existent agent & 0/100 & 0/100\\
\addlinespace[2pt]\multicolumn{3}{l}{\textit{G2 Attribution}}\\
A5\;Impersonate the reviewer to the gateway & 100/100 & 0/100\\
A9\;Replay a captured valid request & 100/100 & 0/100\\
A12\;Publish with no credentials & n/a$^{a}$ & 0/100\\
\addlinespace[2pt]\multicolumn{3}{l}{\textit{G3 Non-interference}}\\
A6\;Read the gateway's mailbox & 100/100 & 0/100\\
A7\;Acknowledge (consume) the gateway's messages & 100/100 & 0/100\\
\addlinespace[2pt]\multicolumn{3}{l}{\textit{G4 Protected oversight}}\\
A8\;Read the full audit ledger & 100/100 & 0/100\\
A10\;Open a connection to the internet & 100/100 & 0/100\\
\bottomrule
\end{tabular}

\smallskip
\footnotesize $^{a}$Not applicable: the baseline relay has no credentials to omit.
\end{table}

Table~\ref{tab:attacks} shows the outcome of all attempts. In the baseline, network isolation and the route policy met the containment requirement (G1), blocking 500 of 500 attempts. Attacks on attribution, non-interference and protected oversight (G2--G4) succeeded in 600 of 600 attempts. For example, the worker impersonated the reviewer in 100 of 100 attempts (95\% CI 96--100\%). The hardened variant blocked every one of the 1{,}200 attempts, including the 100 attempts to publish without credentials; with no internet route, connection attempts to the internet failed as well.

\paragraph{Audit ledger.}

Every relayed impersonation attempt was logged in both variants (100 baseline, 100 hardened). In the baseline, 100 of 100 of these entries named the reviewer as the sender and none named the worker: the ledger recorded the forgery as genuine traffic. In the hardened variant 100 of 100 entries carried the worker's authenticated identity. Reads of another agent's mailbox left no ledger entry in the baseline (0 entries), and the 100 hijacked acknowledgements were logged as ordinary acknowledgements, with no record of which agent made them. In the hardened variant every refused mailbox read and acknowledgement was logged with the caller's authenticated identity (100 of 100 and 100 of 100 entries attributed to the worker).

\subsection{Cost of mediation and hardening}

\begin{table}[t]
\centering\small
\caption{Handoff latency in ms, median (95th percentile), and the paired relay-minus-direct median difference with bootstrap 95\% CI. 1{,}000 pairs per row.}
\label{tab:handoff}
\begin{tabular}{llccc}
\toprule
Variant & Payload & Direct & Relay & Relay $-$ direct\\
\midrule
Baseline & 1\,KB & 2.0 (3.4) & 2.2 (4.0) & 0.29 [0.26, 0.33]\\
 & 10\,KB & 2.1 (3.6) & 2.3 (4.1) & 0.29 [0.26, 0.32]\\
 & 100\,KB & 2.6 (4.9) & 2.9 (5.7) & 0.23 [0.19, 0.27]\\
\addlinespace
Hardened & 1\,KB & 2.2 (4.2) & 2.8 (5.0) & 0.53 [0.50, 0.55]\\
 & 10\,KB & 2.4 (3.8) & 2.9 (5.0) & 0.52 [0.46, 0.56]\\
 & 100\,KB & 2.5 (4.2) & 2.9 (5.1) & 0.43 [0.39, 0.46]\\
\bottomrule
\end{tabular}
\end{table}

\begin{figure}[t]\centering
\includegraphics[width=\linewidth]{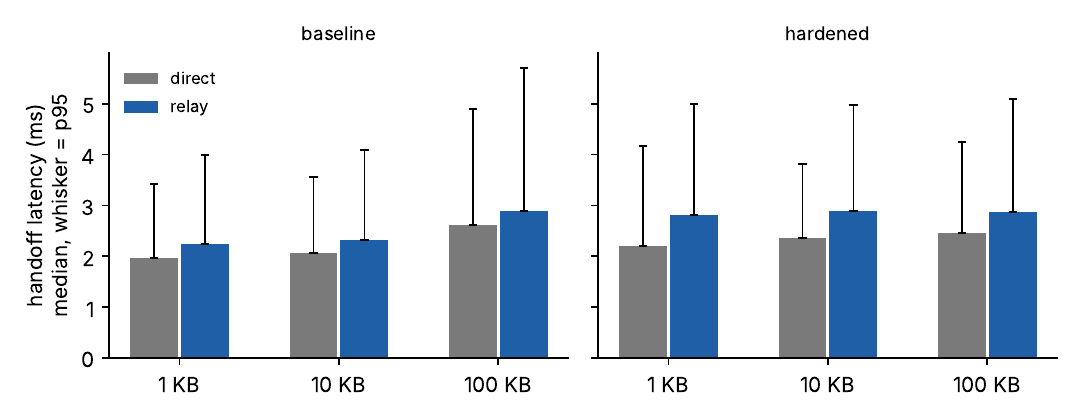}
\caption{Handoff latency by payload size. Bars show medians; whiskers extend to the 95th percentile.}
\label{fig:handoff}\end{figure}

Table~\ref{tab:handoff} shows handoff latency by payload size (Figure~\ref{fig:handoff}). For 1\,KB messages the relay added a median of 0.29\,ms in the baseline (relay slower in 76\% of pairs) and 0.53\,ms in the hardened variant. None of the 12{,}000 handoff requests failed.

\begin{table}[t]
\centering\small
\caption{Throughput (messages/s) and 95th-percentile request latency (ms) with concurrent senders; 400 messages of 1\,KB per cell.}
\label{tab:concurrency}
\begin{tabular}{llcccc}
\toprule
 & & \multicolumn{2}{c}{Throughput} & \multicolumn{2}{c}{p95 latency}\\
\cmidrule(lr){3-4}\cmidrule(lr){5-6}
Variant & Senders & Direct & Relay & Direct & Relay\\
\midrule
Baseline & 1 & 501 & 374 & 2.6 & 3.6\\
 & 4 & 662 & 635 & 8.2 & 9.1\\
 & 16 & 594 & 706 & 45.2 & 47.6\\
\addlinespace
Hardened & 1 & 483 & 310 & 3.0 & 5.2\\
 & 4 & 644 & 540 & 13.3 & 10.8\\
 & 16 & 619 & 651 & 33.3 & 32.4\\
\bottomrule
\end{tabular}
\end{table}

Under concurrency (Table~\ref{tab:concurrency}) relay throughput with 16 senders was 706 messages/s in the baseline and 651 in the hardened variant (hardening reduced it by 8\%). Senders shared the worker container's 0.5-CPU limit, and the relay had twice the direct receiver's CPU allowance, so these figures compare the two relay variants rather than measure either path's ceiling; with 16 senders the relay outpaced the direct route for that reason.

\begin{table}[t]
\centering\small
\caption{End-to-end delivery time (ms) from sending to retrieval by the receiver; 200 messages at 10/s.}
\label{tab:delivery}
\begin{tabular}{llccc}
\toprule
Variant & Path & $n$ & Median & p95\\
\midrule
Baseline & Direct & 200 & 2.0 & 2.7\\
 & Relay, poll 0.05\,s & 200 & 31.0 & 56.6\\
 & Relay, poll 0.5\,s & 200 & 265 & 498\\
\addlinespace
Hardened & Direct & 200 & 2.1 & 3.3\\
 & Relay, poll 0.05\,s & 200 & 32.4 & 57.6\\
 & Relay, poll 0.5\,s & 200 & 267 & 496\\
\bottomrule
\end{tabular}
\end{table}

End to end (Table~\ref{tab:delivery}), delivery through the relay was dominated by the receiver's polling interval: median 265\,ms at the agents' 0.5\,s interval and 31.0\,ms at 0.05\,s, against 2.0\,ms for a direct send.

\section{Discussion}
\paragraph{Containment is necessary but not sufficient.} The baseline did what containment systems are built to do: no attempt to bypass the relay or break routing policy succeeded in either variant. That is the property container-based agent sandboxes usually verify, and Connexin's own test suite verified it. Governance needs more. The relay trusted the sender name written in each message, served any mailbox to any caller and exposed its ledger to every agent, so a compromised worker could speak as the reviewer, read and silently consume the gateway's results and read every agent's activity. None of these attacks crosses a network boundary, so none is visible to network isolation, and none was caught by a verification suite that tested intended behaviour.

\paragraph{An audit trail that looks authoritative but is wrong.} The finding most relevant to clinical governance is that the ledger recorded forged actions as genuine. Every impersonation attempt appeared as a normal message from the reviewer, hijacked acknowledgements appeared as normal acknowledgements with no record of who made them, and reads of another agent's mailbox left no trace at all. An incident review based on this ledger would have attributed the worker's actions to the reviewer, and so to the reviewer's model, prompt and data. A log that is trusted but wrong is arguably worse than a missing one, because it leads reviewers to confident, incorrect conclusions. Record-keeping obligations such as those in the EU AI Act~\cite{euaiact2024} implicitly assume that what is logged is true; our results show that this has to be engineered, not assumed. The hardened relay logged the authenticated identity beside the claimed one, so every refused relay request that carried the worker's credential was attributed to the worker, and requests without a credential were logged as unauthenticated.

\paragraph{Cost.} \DiscussionCost{}

\paragraph{Questions for governance review.} Committees approving multi-agent AI systems for clinical research can turn these findings into review questions:
\begin{enumerate}
\item Where does each agent's identity come from? It should come from credentials issued by the mediator, never from message content.
\item Does the audit log record the authenticated identity of every action, including refused ones?
\item Can one agent read, acknowledge or otherwise interfere with messages meant for another?
\item Who can read the audit log? Agents should not.
\item Can agents reach the internet or other outside systems directly?
\item Has the system been tested with a compromised agent, not only with functional checks?
\end{enumerate}

\paragraph{Limitations.}
\begin{itemize}
\item One host, one relay instance and two agents, with synthetic payloads and no LLM in the loop. The attacker was scripted, so the results bound what a compromised agent \emph{can} do, not what an LLM would choose to do.
\item The relay had twice the CPU allowance of the direct receiver (1 versus 0.5 CPU), as deployed. This matters little for one sender at a time but favours the relay under concurrency, so throughput should be compared between variants rather than between relay and direct routes.
\item Other workloads shared the host. Both variants ran under the same conditions and handoff measurements were paired, but absolute latencies are specific to this host.
\item Bearer tokens on an internal network protect against a compromised peer, not against an attacker who can read relay traffic or memory; mutual TLS or signed messages, with key rotation, would be needed for that.
\item Removing the internet route stops direct exfiltration, but an agent can still put data into messages it is allowed to send. Content-level controls are outside our scope.
\item We did not test container escape~\cite{marchand2026sandboxescape}, attacks on the relay itself, or the governance processes that would act on the audit trail.
\end{itemize}
The last point is the natural next step. With an attributable audit trail in place, the same environment can be used to simulate clinical governance itself: to test which oversight designs detect a failing or subverted agent fastest.

\section{Conclusion}
A standard containment design kept a compromised agent within its routes but let it speak as another agent, read other agents' work and corrupt the record of who did what. For clinical governance, which depends on that record, this is a fundamental gap. Taking identity from credentials, scoping access to owners and protecting the audit log closed the gaps we tested, at low cost. Verifiable agent identity and an attributable audit trail should be treated as baseline requirements for multi-agent LLM systems in clinical research, not as optional hardening.

\section*{Funding}
This work was supported by the Saudi National Institute of Health (Saudi NIH) under Grant PRI01-2401-KAU19-46641119.

\section*{Author contributions}
Following the CRediT taxonomy. \textbf{Anas H. Alzahrani:} conceptualization, methodology, investigation, formal analysis, writing (original draft), supervision. \textbf{Omar F. Alaskar:} software, system design, investigation, writing (review and editing). Both authors approved the final manuscript.

\section*{Code and data availability}\label{sec:availability}
All code, including the baseline and hardened relays, the experiment kit and the raw results, is available at \url{https://github.com/AnasAlzahrani/connexin-paper}.

\section*{Use of AI tools}
Parts of the Connexin codebase were written with the OpenAI Codex coding agent under the authors' direction. The experiment kit, the hardened relay variant and drafts of this manuscript were prepared with assistance from Anthropic's Claude. The authors reviewed all code, results and text and take full responsibility for them.

\bibliographystyle{unsrtnat}
\bibliography{references}

\end{document}